\documentclass[aps,pra,preprint,superscriptaddress,longbibliography]{revtex4-2}

\usepackage{amsmath}
\usepackage{amsthm}
\usepackage{amsfonts}
\usepackage{amssymb}
\usepackage[dvipsnames]{xcolor}
\usepackage{graphicx}
\usepackage{tikz}
\usepackage{color}
\usepackage[pdftex]{hyperref} 
\usepackage{longtable}
\usepackage{array}
\usepackage{dsfont}

\begin{document}
	
	\title{Orbital Angular Momentum Coherent State Beams}

        \author{D. Aguirre-Olivas}
        \email[e-mail: ]{dilia@ifc.unam.mx}
        \affiliation{Instituto de Fisiología Celular - Neurociencias, Universidad Nacional Autónoma de México, Ciudad de México, 04510, Mexico}

        \author{G. Mellado-Villase\~nor}
        \email[e-mail: ]{gmelladov@fisica.unam.mx}
        \affiliation{Instituto de Física, Universidad Nacional Autónoma de México, Ciudad de México, 04510, Mexico}

        \author{B. Perez-Garcia}
        \email[e-mail: ]{b.pegar@tec.mx}
        \affiliation{Photonics and Mathematical Optics Group, Tecnologico de Monterrey, Monterrey 64849, Mexico} 
        
        \author{B.~M. Rodr\'iguez-Lara}
        \email[e-mail: ]{bmlara@upp.edu.mx; blas.rodriguez@gmail.com}
        \affiliation{Universidad Polit\'ecnica de Pachuca. Carr. Pachuca-Cd. Sahag\'un Km.20, Ex-Hda. Santa B\'arbara. Zempoala, 43830 Hidalgo, Mexico }
	
	\date{\today}

\begin{abstract} 
We explore a family of paraxial beams constructed by the linear superposition of Laguerre–Gaussian beams, representing an optical analogue to generalized $SU(2)$ Lie group coherent states.
A single complex parameter controls a smooth transition between Laguerre–Gaussian and Hermite–Gaussian beams, with intermediate beams that merge characteristics of both families. 
Our beams exhibit propagation-invariant properties, up to a scaling factor, a highly desirable feature for optical applications, validated via holographic experimental results. 
\end{abstract}

\maketitle
\newpage	


Paraxial beams (PBs), solutions to the paraxial wave equation (PWE) \cite{Lax1975b}, have garnered significant attention due to their unique properties and wide-ranging applications \cite{Kiselev2007, Forbes2021}.
The fundamental families of PBs in linear, isotropic, and homogeneous media are Hermite–Gaussian beams (HGBs) and Laguerre–Gaussian beams (LGBs) \cite{Siegman1986lasers}. 
HGBs are defined by horizontal and vertical numbers in Cartesian coordinates, corresponding to nodes in those directions and associated to linear momentum. 
LGBs are defined in cylindrical coordinates, with radial and azimuthal numbers corresponding to nodes in those directions and associated with intrinsic hyperbolic momentum and orbital angular momentum (OAM) \cite{Allen1992, Karimi2007, Plick2015}.
Both HGBs and LGBs form orthogonal bases that retain their transverse shapes, up to a scaling factor, during propagation.

These properties make HGBs and LGBs essential tools in fields such as optical communication \citep{Wang2022}, microscopy \cite{Liu2007,Mur2013}, information processing \citep{Fontaine2019}, and beam shaping \citep{Droop2021}. 
A promising approach to expanding their applications involves their coherent or incoherent superposition, allowing control over intensity and phase \cite{Parisi2014, Volyar2022}. 
Building upon this approach, recent studies demonstrated optical analogues to quantum coherent states \cite{Morales2024a} and generalized coherent states \cite{Morales2024b}, producing structured light with controllable intensity and phase. 
These analogies, when propagated through parabolic graded-index (GRIN) fibers, exhibit interesting behaviors such as harmonic motion in the form of accelerating or breathing beams \cite{Collado2024}, and can be extended to displaced, rotated, and squeezed number states \cite{Morales2024b}.

Three well-known procedures bridge HGBs and LGBs. 
Ince–Gaussian beams (IGBs), solutions to the PWE in elliptical coordinates, produce a continuous transition by adjusting the ellipticity parameter \cite{Bandres2004}. 
Hermite–Laguerre–Gaussian beams (HLGBs) \cite{Abramochkin2004}, derived from an astigmatic transformation of HGBs \cite{Abramochkin1991}, also link HGBs and LGBs using a single parameter. 
Generalized Hermite–Gaussian beams (gHGBs) \cite{Wang2016} involve the superposition of HGBs, with a parameter controlling the beam shape. 
When this parameter equals an integer multiple of $\pi/2$, the HLGBs are recovered. 
In both HLGBs and gHGBs, the beam shape is determined by a single real parameter.

In this Letter, we propose a simple and intuitive method to connect HGBs and LGBs, based on their inherent symmetries. 
At the plane $z = 0$, these beams reduce to Hermite-Gauss modes (HGMs) and Laguerre-Gauss modes (LGMs), which satisfy a scaled Schrödinger equation for a two-dimensional isotropic harmonic oscillator. 
This oscillator exhibits various symmetries \cite{Morales2024a, Morales2024b}, one of which is the two-mode special unitary Lie group $SU(2)$, associated with rotations on the Bloch-Poincar\'e sphere for spin-$1/2$ systems. 
Using this symmetry, we construct a paraxial optics analogue of generalized orbital angular momentum coherent states in terms of a complex coherent parameter $\alpha$. 
Applying this rotation to an LGB with real $\alpha = \pi/4$ transforms it into an HGB.
The complex coherent parameter phase rotates the intensity and phase distribution around the optical axis. 
Our method provides a general and intuitive framework for bridging HGBs and LGBs, creating a broader family of paraxial beams with propagation-invariant properties, up to a scaling factor.


We define scalar orthonormal LGBs,
\begin{align}
    \Psi_{p,\ell}(r,\theta,z) = \frac{\sqrt{2}}{w(z)} e^{-\frac{i k r^2}{2 R(z)}} e^{ i (2 p + \vert \ell \vert +1) \varphi(z)} \psi_{p,\ell} \left( \rho, \theta \right), \label{eq: LGBs}
\end{align}
in terms of the standard Gaussian beam parameters; the beam width $w(z) = w_{0} \sqrt{1 + (z/z_{R})^{2}}$, curvature radius $R(z) = (z^{2} + z_{R})/z$, Gouy phase $\varphi(z) = \tan^{-1}(z/z_{R})$, waist radius $w_{0} = \sqrt{\lambda z_{R}/\pi}$, Rayleigh range $z_{R} = \pi w_{0}^{2}/\lambda$, wavenumber $k = 2 \pi/\lambda$, wavelength $\lambda$. It is also important to mention that Eq.\,(\ref{eq: LGBs}) has units of meters$^{-1}$ due to orthonormalization process. 
To properly construct optical fields, a constant with units of volts should be included.
We define a scaled variable $\rho = \sqrt{2}r/w(z)$ for the sake of space.
They satisfy the PWE,
\begin{align}
  \left(\nabla_{\perp}^{2} - 2 i k \frac{d}{dz} \right) \Psi_{p,\ell}(r, \theta, z) = 0,
\end{align}
in circular-cylinder coordinates.
The radial number $p$ and azimuthal numbers $\ell$ characterize the beam by determining the number of nodes in the corresponding directions.
Our LGBs form a complete orthonormal basis for constructing any light field distribution in the transverse plane at any given propagation distance $z$. 

To draw an analogy with quantum mechanics, we use the quantum Laguerre-Gaussian modes (LGMs),
\begin{align}
    \psi_{p, \ell}(\rho, \phi) 
    = (-1)^{p} \sqrt{\frac{p!}{\pi (p+\vert \ell \vert)!}} \, \rho^{\vert \ell \vert} e^{ - \frac{1}{2} \rho^{2}} \mathrm{L}_{p}^{\vert \ell \vert} (\rho^{2}) e^{i \ell \phi},
\end{align}
expressed in terms of the generalized Laguerre polynomials $L_{p}^{\vert \ell \vert}(\rho^{2})$.
They solve the isotropic two-dimensional quantum harmonic oscillator,
\begin{align}
        \frac{1}{2}\left(-\nabla_{\perp}^{2} + \rho^{2}\right)\psi_{p, \ell}(\rho, \phi) = (2 p + \vert \ell \vert + 1) \psi_{p, \ell}(\rho, \phi),
\end{align}
in dimensionless polar coordinates $(\rho, \phi)$.
In quantum mechanics, LGMs provide a complete orthonormal basis for a Hilbert space with diverse underlying symmetries.
These symmetries enable the construction of distinct families of generalized coherent states \cite{Morales2024a, Morales2024b}.
Generalized coherent states of LGMs emerge from subspaces defined by conserved variables, are normalized, and provide overcomplete bases for their respective subspaces.

We construct a paraxial optics analogue to generalized coherent states of the $SU(2)$ Lie group,
\begin{align}
    \Psi_{j,m,\alpha} (\rho,\theta,z) = \sum_{q=0}^{2 j} c_{q}(j,m,\alpha) \Psi_{j- \vert j-q \vert, 2 (j-q)}(\rho,\theta,z) \,,
    \label{eq:lsupLGBs}
\end{align}
where we derive the weight coefficients,
\begin{align}
    \begin{aligned}
        c_{q}(j, m, \alpha) &= \sqrt{\binom{2j}{q} \binom{2j}{j-m}} \, \times \\
        & \times \,_{2}\mathrm{F}_{1}\left(-q,-j+m; -2j; \csc^{2} \vert \alpha \vert\right) \, \times \\
        & \times e^{i \phi (j-q-m)} \left(\sec \vert \alpha \vert\right)^{-2j} \left(\tan \vert \alpha \vert\right)^{j+q-m},
    \end{aligned}
    \label{eq:coeffs}
\end{align}
from the spatial light mode analogue of the two-dimensional harmonic oscillator quantum scenario \cite{Morales2024a, Morales2024b}.
The complex coherent parameter $\alpha = \vert \alpha \vert e^{i \phi}$ controls the beam properties.
Its amplitude $\vert \alpha \vert$ and phase $\phi$ are real-valued and exhibit periodicities of $\pi$ and $2\pi$, respectively. 
The indices $j$ and $m$ are non-negative integers constrained by $ m  \leq j$.
We use the binomial coefficient $\binom{a}{b} = a!/[b!(a-b)!]$ and the Gauss hypergeometric function ${}_{2}\mathrm{F}_{1}(a,b;c;z)$. 
In our optical analogy, the Gouy phase from the Gaussian beams plays a key role, influencing both the intensity and phase distributions based on the  radial and azimuthal numbers and the propagation distance, as we observe in Eq. \ref{eq: LGBs}. From this, it is noteworthy that the terms in the summation of Eq. \ref{eq:lsupLGBs} are independent of the Gouy phase, as they all share the same Gouy phase factor, \(\left(2p+\vert l \vert +1 \right)\phi(z) = (2j+1)\phi(z)\). 
Hereby, we will refer to our beams as orbital angular momentum coherent state (OAMCS) beams.

\begin{figure}[htp!]
    \centering    \includegraphics[width=\textwidth]{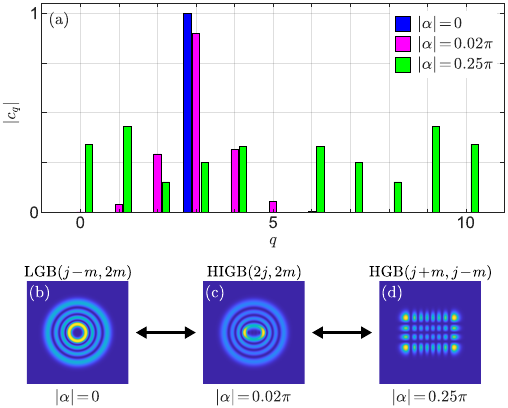}
    \caption{(a) Bar graph of $\vert c_q \vert$ versus $q$ for parameters $\left\{j,m,\phi\right\}\!=\!\left\{5,2,0\right\}$ and three values of $\vert\alpha\vert$. (b)-(d): Corresponding OAMCS intensity profiles.}
    \label{fig:Cq}
\end{figure}

Let us explore our OAMCS beams for real coherent parameters, $\phi = 0$.
A zero coherent parameter yields a weight coefficient $c_{q}(j,m,\alpha=0) = \delta_{q,j-m}$, Fig. \ref{fig:Cq}(a), producing a LGB with radial number $p = j - \vert m \vert$ and azimuthal number $\ell = 2 m$, Fig. \ref{fig:Cq}(b).
A small increasing in the amplitude, for example $\vert \alpha \vert = 0.02 \pi$ in Fig. \ref{fig:Cq}(a), distributes the weight $q = j-m$, producing an elliptical deformation of the LGB, reminiscent of Helical Ince-Gaussian beams (HIGBs) \cite{Bandres2004}, Fig. \ref{fig:Cq}(c).
As the amplitude increases further, more nonzero weight coefficients appear, producing a larger elliptical deformation  until the amplitude reaches the value $\vert \alpha \vert = \pi / 4$, where we recover a HGB with horizontal number $j+m$ and vertical number $j-m$, Fig. \ref{fig:Cq}(d).

\begin{figure}[htp]
    \centering
    \includegraphics[width=\textwidth]{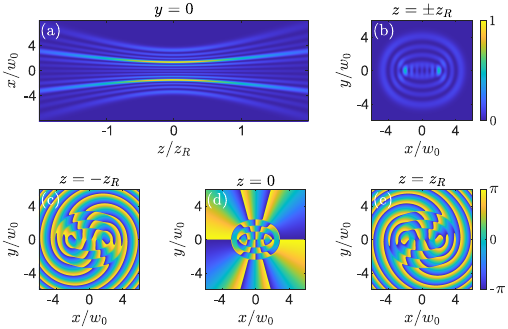}
    \caption{(a) Longitudinal at $y=0$ and (b) transverse at $z= \pm z_{R}$ for our OAMCS beam with parameters $\{ j , m , \vert\alpha\vert, \phi \} = \{6, 3, \pi/16, 0 \}$.
    Phase patterns at the transverse planes (c) $z=-z_R$, (d) $z=0$, and (e) $z=z_R$.}
    \label{fig:zL}
\end{figure}

Figure \ref{fig:zL} shows the intensity and phase distributions of our OAMCS with parameters $\{ j , m , \vert\alpha\vert, \phi \} = \{6, 3, \pi/16, 0 \}$. 
The longitudinal intensity pattern at $y=0$, Fig. \ref{fig:zL}(a), reveals a symmetry around $z=0$ that remains preserved during propagation, Fig. \ref{fig:zL}(b), characteristic of spatial invariant beams. 
In the transverse planes at $z=\pm z_{R}$, Figs. \ref{fig:zL}(c) and \ref{fig:zL}(e), the phase distributions exhibit identical patterns with opposite signs. 
These phase distributions show spiral structural changes due to both the spherical and Gouy phases, in contrast to the phase pattern in the plane at $z=0$, Fig. \ref{fig:zL}(d).

 \newpage

\begin{figure}[htp]
    \centering
    \includegraphics[width=\textwidth]{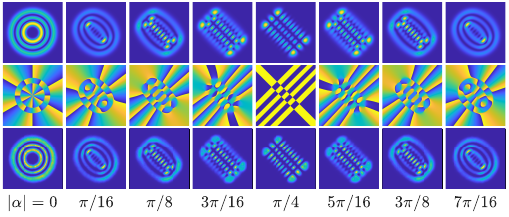}
    \caption{Theoretical intensity (first row) and phase distribution (second row) at the plane $z=0$, along with their corresponding experimental intensities (third row) for the beam in \eqref{eq:lsupLGBs} with parameters $\{j, m, \phi \}=\{ 5, 3, \pi/2 \}$, and variable real coherent parameter $\vert\alpha\vert \in [0, \pi/2 )$. For the experimental results, we set $w_0 = 0.4$ mm and $\lambda = 632.8$ nm.}
    \label{fig:alpha}
\end{figure}

The complex coherent parameter amplitude $\vert \alpha \vert$ controls a smooth transition from LGBs with radial number $p = j - \vert m \vert$ and azimuthal number $\ell = 2 m$ at $\vert\alpha\vert = 0$, to HGBs with horizontal number $n_{x} = j + m = (2 p + \vert \ell \vert + \ell)/2$ and vertical number $n_{y} = j - m = (2 p + \vert \ell \vert - \ell )/2 $ at $\vert\alpha\vert = \pi/4$, and back to LGBs with $p = j - \vert m \vert$ and $\ell = -2 m$ at $\vert\alpha\vert = \pi /2$. 
For amplitude values near zero and $\pi/2$, our OAMCS exhibit elliptical symmetry reminiscent of HIGBs, gradually transitioning to rectangular symmetry as the amplitude approaches $\pi/4$, then returning to elliptical symmetry for values close to $\pi/2$.
Figure \ref{fig:alpha} illustrates this transition, showing the intensity (first row) and phase (second row) distributions for a real coherent parameter in the range $\vert \alpha \vert \in [0, 7 \pi / 16]$ with $\phi=\pi/2$.
The coherent parameter amplitude shows $\pi$-periodicity, recovering LGBs at $\vert \alpha \vert = n \pi /2$ and HGBs for $\vert \alpha \vert = (2 n + 1) \pi /2$ with $n = 0, 1, 2, \ldots$
Intermediate amplitude values produce OAMCS that combine the characteristics of both LGBs and HGBs. Experimental results for the intensity are shown in the third row.

\begin{figure}[htp]
    \centering    \includegraphics[width=\textwidth]{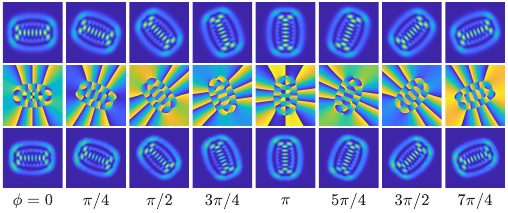}
    \caption{Theoretical intensity (first row) and phase distribution (second row) at the plane $z=0$, along with their corresponding experimental intensities (third row) for the beam in \eqref{eq:lsupLGBs} with parameters $\{j, m, \vert \alpha \vert \} = \{ 5, 3, \pi/8\}$, and variable coherent parameter phase $\phi \in [0, 2 \pi )$. For the experimental results, we set $w_0 = 0.4$ mm and $\lambda = 632.8$ nm.}
    \label{fig:phi}
\end{figure}

The complex coherent parameter phase $\phi$ rotates the intensity and phase distributions clockwise by an angle $\phi/2$ around the optical axis, while keeping a fixed coherent parameter amplitude. 
Figure \ref{fig:phi} illustrates this rotation for our OAMCS, intensity (first row) and phase (second row), with parameters $\{j, m, \vert \alpha \vert \}=\{ 5, 3, \pi/8\}$ in a phase range $\phi \in [ 0, 7 \pi / 4]$. 
The complex coherent parameter phase exhibits $2\pi$-periodicity. Experimental results for the intensity are depicted in the third row.

\begin{figure}[htp]
    \centering    \includegraphics[width=\textwidth]{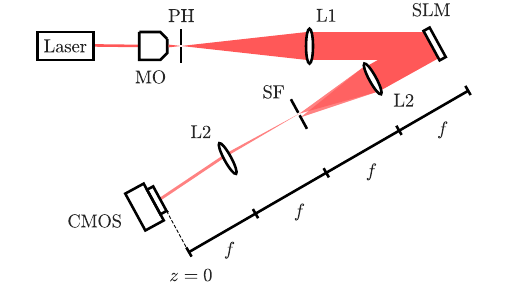}
    \caption{Experimental setup.  Laser, He-Ne source;  MO, Microscope Objective;  PH, Pinhole; L1--L2, lenses; SLM, Spatial Light Modulator; SF, Spatial Filter; CMOS, sensor.}
    \label{fig:setup}
\end{figure}

Previously, LGBs, HGBs, and IGBs have been efficiently generated using diffractive phase elements \cite{Matsumoto2008, Aguirre2015}. 
However, these techniques are tailored to specific beam types and are not suited for generating any given sub-families. 
To generate our paraxial OAMCS beams, we used a synthetic phase hologram (SPH) \cite{Arrizon2007}, a holographic technique capable of producing optical fields with arbitrary amplitude and phase.
In our experiment (Fig.\ \ref{fig:setup}), we expanded and collimated a linearly polarized He-Ne laser beam $(\lambda = 632.8 \text{ nm})$ to illuminate a phase-only spatial light modulator (SLM, Holoeye PLUTO VIS), which is programmed with a SPH encoding the complex amplitudes of our OAMCS. 
The beam, encoded in the first diffraction order of the SPH, is processed through a $4f$ spatial filtering system and captured using a CMOS sensor (Nikon D7500).

\begin{figure}[htp]
    \centering    \includegraphics[width=\textwidth]{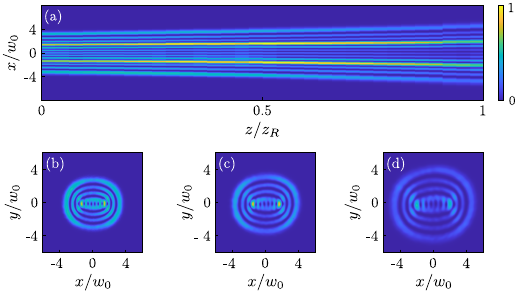}
    \caption{Experimental propagation of an OAMCS beam with parameters $\{ j , m , \vert\alpha\vert, \phi \} = \{6, 3, \pi/16, 0 \}$ and $w_0 = 0.4$ mm. (a) Shows the intensity at $y=0$ cut of the longitudinal propagation. (b)--(d) Depicts the transverse intensity at the planes (b) $z=0$, (c) $z=z_R/2$ and (d) $z=z_R$, respectively.}
    \label{fig:expprop}
\end{figure}

Figure \ref{fig:expprop} presents the experimental results for the propagation of an OAMCS beam with parameters $\{ j , m , \vert \alpha \vert \} = \{6, 3, \pi/16 \}$ and $w_0 = 0.4$ mm.  Fig.\ \ref{fig:expprop}(a) shows the experimental longitudinal propagation at $y=0$, measured by moving the CMOS sensor along the $z$--axis in increments of $z_R/32$.  Figures \ref{fig:expprop}(b)--(d), display the transverse intensity at planes $z=0$,  $0.5z_R$ and $z_R$, respectively. Note that the beam features shape--invariant propagation, aside from a scaling factor. These results align with the theoretical findings depicted in Fig. \ref{fig:zL}. 

In conclusion, we constructed a paraxial optics analogue to generalized coherent states of the $SU(2)$ Lie group.
Our orbital angular momentum coherent state beams are a family of paraxial beams generated by a $SU(2)$ rotation, defined by a complex coherent parameter, acting on any given LGB.
They take the form of a coherent superposition of LGBs weighted by coefficients recovered from the Lie algebra framework.
For a real complex coherent parameter, our OAMCS transition smoothly from LGBs to HGBs, with intermediate beams merging features of both families. 
The complex parameter phase rotates their intensity and phase distributions around the optical axis. 
Our symmetry-based framework expands the family of propagation-invariant structured beams, offering new avenues for the manipulation of scalar light beams.

\section*{Funding}
Not Applicable.

\section*{Acknowledgments}
Not Applicable.

\section*{Disclosures}
The authors declare no conflicts of interest.

\section*{Data Availability Statement}
All the data are available from the corresponding author upon reasonable request.




\begin{thebibliography}{24}%
\makeatletter
\providecommand \@ifxundefined [1]{%
 \@ifx{#1\undefined}
}%
\providecommand \@ifnum [1]{%
 \ifnum #1\expandafter \@firstoftwo
 \else \expandafter \@secondoftwo
 \fi
}%
\providecommand \@ifx [1]{%
 \ifx #1\expandafter \@firstoftwo
 \else \expandafter \@secondoftwo
 \fi
}%
\providecommand \natexlab [1]{#1}%
\providecommand \enquote  [1]{``#1''}%
\providecommand \bibnamefont  [1]{#1}%
\providecommand \bibfnamefont [1]{#1}%
\providecommand \citenamefont [1]{#1}%
\providecommand \href@noop [0]{\@secondoftwo}%
\providecommand \href [0]{\begingroup \@sanitize@url \@href}%
\providecommand \@href[1]{\@@startlink{#1}\@@href}%
\providecommand \@@href[1]{\endgroup#1\@@endlink}%
\providecommand \@sanitize@url [0]{\catcode `\\12\catcode `\$12\catcode `\&12\catcode `\#12\catcode `\^12\catcode `\_12\catcode `\%12\relax}%
\providecommand \@@startlink[1]{}%
\providecommand \@@endlink[0]{}%
\providecommand \url  [0]{\begingroup\@sanitize@url \@url }%
\providecommand \@url [1]{\endgroup\@href {#1}{\urlprefix }}%
\providecommand \urlprefix  [0]{URL }%
\providecommand \Eprint [0]{\href }%
\providecommand \doibase [0]{https://doi.org/}%
\providecommand \selectlanguage [0]{\@gobble}%
\providecommand \bibinfo  [0]{\@secondoftwo}%
\providecommand \bibfield  [0]{\@secondoftwo}%
\providecommand \translation [1]{[#1]}%
\providecommand \BibitemOpen [0]{}%
\providecommand \bibitemStop [0]{}%
\providecommand \bibitemNoStop [0]{.\EOS\space}%
\providecommand \EOS [0]{\spacefactor3000\relax}%
\providecommand \BibitemShut  [1]{\csname bibitem#1\endcsname}%
\let\auto@bib@innerbib\@empty
\bibitem [{\citenamefont {W\"unsche}(1989)}]{Lax1975b}%
  \BibitemOpen
  \bibfield  {author} {\bibinfo {author} {\bibfnamefont {A.}~\bibnamefont {W\"unsche}},\ }\bibfield  {title} {\bibinfo {title} {Generalized {Gaussian} beam solutions of paraxial optics and their connection to hidden symmetry},\ }\href {https://doi.org/10.1364/JOSAA.6.001320} {\bibfield  {journal} {\bibinfo  {journal} {J. Opt. Soc. Am. A}\ }\textbf {\bibinfo {volume} {6}},\ \bibinfo {pages} {1320} (\bibinfo {year} {1989})}\BibitemShut {NoStop}%
\bibitem [{\citenamefont {Kiselev}(2007)}]{Kiselev2007}%
  \BibitemOpen
  \bibfield  {author} {\bibinfo {author} {\bibfnamefont {A.~P.}\ \bibnamefont {Kiselev}},\ }\bibfield  {title} {\bibinfo {title} {Localized light waves: Paraxial and exact solutions of the wave equation (a review)},\ }\href {https://doi.org/10.1134/S0030400X07040200} {\bibfield  {journal} {\bibinfo  {journal} {Opt. Spectrosc.}\ }\textbf {\bibinfo {volume} {102}},\ \bibinfo {pages} {603} (\bibinfo {year} {2007})}\BibitemShut {NoStop}%
\bibitem [{\citenamefont {Forbes}\ \emph {et~al.}(2021)\citenamefont {Forbes}, \citenamefont {{de Oliveira}},\ and\ \citenamefont {Dennis}}]{Forbes2021}%
  \BibitemOpen
  \bibfield  {author} {\bibinfo {author} {\bibfnamefont {A.}~\bibnamefont {Forbes}}, \bibinfo {author} {\bibfnamefont {M.}~\bibnamefont {{de Oliveira}}},\ and\ \bibinfo {author} {\bibfnamefont {M.~R.}\ \bibnamefont {Dennis}},\ }\bibfield  {title} {\bibinfo {title} {Structured light},\ }\href {https://doi.org/10.1038/s41566-021-00780-4} {\bibfield  {journal} {\bibinfo  {journal} {Nat. Photonics}\ }\textbf {\bibinfo {volume} {15}},\ \bibinfo {pages} {253} (\bibinfo {year} {2021})}\BibitemShut {NoStop}%
\bibitem [{\citenamefont {Siegman}(1986)}]{Siegman1986lasers}%
  \BibitemOpen
  \bibfield  {author} {\bibinfo {author} {\bibfnamefont {A.~E.}\ \bibnamefont {Siegman}},\ }\href@noop {} {\emph {\bibinfo {title} {Lasers}}}\ (\bibinfo  {publisher} {University Science Books},\ \bibinfo {address} {Mill Valley, Calif},\ \bibinfo {year} {1986})\BibitemShut {NoStop}%
\bibitem [{\citenamefont {Allen}\ \emph {et~al.}(1992)\citenamefont {Allen}, \citenamefont {Beijersbergen}, \citenamefont {Spreeuw},\ and\ \citenamefont {Woerdman}}]{Allen1992}%
  \BibitemOpen
  \bibfield  {author} {\bibinfo {author} {\bibfnamefont {L.}~\bibnamefont {Allen}}, \bibinfo {author} {\bibfnamefont {M.~W.}\ \bibnamefont {Beijersbergen}}, \bibinfo {author} {\bibfnamefont {R.~J.~C.}\ \bibnamefont {Spreeuw}},\ and\ \bibinfo {author} {\bibfnamefont {J.~P.}\ \bibnamefont {Woerdman}},\ }\bibfield  {title} {\bibinfo {title} {Orbital angular momentum of light and the transformation of {Laguerre-Gaussian} laser modes},\ }\href {https://doi.org/10.1103/PhysRevA.45.8185} {\bibfield  {journal} {\bibinfo  {journal} {Phys. Rev. A}\ }\textbf {\bibinfo {volume} {45}},\ \bibinfo {pages} {8185} (\bibinfo {year} {1992})}\BibitemShut {NoStop}%
\bibitem [{\citenamefont {Karimi}\ \emph {et~al.}(2007)\citenamefont {Karimi}, \citenamefont {Zito}, \citenamefont {Piccirillo}, \citenamefont {Marrucci},\ and\ \citenamefont {Santamato}}]{Karimi2007}%
  \BibitemOpen
  \bibfield  {author} {\bibinfo {author} {\bibfnamefont {E.}~\bibnamefont {Karimi}}, \bibinfo {author} {\bibfnamefont {G.}~\bibnamefont {Zito}}, \bibinfo {author} {\bibfnamefont {B.}~\bibnamefont {Piccirillo}}, \bibinfo {author} {\bibfnamefont {L.}~\bibnamefont {Marrucci}},\ and\ \bibinfo {author} {\bibfnamefont {E.}~\bibnamefont {Santamato}},\ }\bibfield  {title} {\bibinfo {title} {{Hypergeometric-Gaussian modes}},\ }\href {https://doi.org/10.1364/OL.32.003053} {\bibfield  {journal} {\bibinfo  {journal} {Opt. Lett.}\ }\textbf {\bibinfo {volume} {32}},\ \bibinfo {pages} {3053} (\bibinfo {year} {2007})}\BibitemShut {NoStop}%
\bibitem [{\citenamefont {Plick}\ and\ \citenamefont {Krenn}(2015)}]{Plick2015}%
  \BibitemOpen
  \bibfield  {author} {\bibinfo {author} {\bibfnamefont {W.~N.}\ \bibnamefont {Plick}}\ and\ \bibinfo {author} {\bibfnamefont {M.}~\bibnamefont {Krenn}},\ }\bibfield  {title} {\bibinfo {title} {Physical meaning of the radial index of {Laguerre-Gauss} beams},\ }\href {https://doi.org/10.1103/PhysRevA.92.063841} {\bibfield  {journal} {\bibinfo  {journal} {Phys. Rev. A}\ }\textbf {\bibinfo {volume} {92}},\ \bibinfo {pages} {063841} (\bibinfo {year} {2015})}\BibitemShut {NoStop}%
\bibitem [{\citenamefont {Wang}\ \emph {et~al.}(2022)\citenamefont {Wang}, \citenamefont {Liu}, \citenamefont {Li}, \citenamefont {Zhao}, \citenamefont {Du},\ and\ \citenamefont {Zhu}}]{Wang2022}%
  \BibitemOpen
  \bibfield  {author} {\bibinfo {author} {\bibfnamefont {J.}~\bibnamefont {Wang}}, \bibinfo {author} {\bibfnamefont {J.}~\bibnamefont {Liu}}, \bibinfo {author} {\bibfnamefont {S.}~\bibnamefont {Li}}, \bibinfo {author} {\bibfnamefont {Y.}~\bibnamefont {Zhao}}, \bibinfo {author} {\bibfnamefont {J.}~\bibnamefont {Du}},\ and\ \bibinfo {author} {\bibfnamefont {L.}~\bibnamefont {Zhu}},\ }\bibfield  {title} {\bibinfo {title} {Orbital angular momentum and beyond in free-space optical communications},\ }\href {https://doi.org/doi:10.1515/nanoph-2021-0527} {\bibfield  {journal} {\bibinfo  {journal} {Nanophotonics}\ }\textbf {\bibinfo {volume} {11}},\ \bibinfo {pages} {645} (\bibinfo {year} {2022})}\BibitemShut {NoStop}%
\bibitem [{\citenamefont {Liu}\ and\ \citenamefont {Kim}(2007)}]{Liu2007}%
  \BibitemOpen
  \bibfield  {author} {\bibinfo {author} {\bibfnamefont {C.}~\bibnamefont {Liu}}\ and\ \bibinfo {author} {\bibfnamefont {D.~Y.}\ \bibnamefont {Kim}},\ }\bibfield  {title} {\bibinfo {title} {Differential imaging in coherent {anti-Stokes Raman} scattering microscopy with {Laguerre-Gaussian} excitation beams},\ }\href {https://doi.org/10.1364/OE.15.010123} {\bibfield  {journal} {\bibinfo  {journal} {Opt. Express}\ }\textbf {\bibinfo {volume} {15}},\ \bibinfo {pages} {10123} (\bibinfo {year} {2007})}\BibitemShut {NoStop}%
\bibitem [{\citenamefont {Mur}\ \emph {et~al.}(2013)\citenamefont {Mur}, \citenamefont {Kav\v{c}i\v{c}},\ and\ \citenamefont {Poberaj}}]{Mur2013}%
  \BibitemOpen
  \bibfield  {author} {\bibinfo {author} {\bibfnamefont {J.}~\bibnamefont {Mur}}, \bibinfo {author} {\bibfnamefont {B.}~\bibnamefont {Kav\v{c}i\v{c}}},\ and\ \bibinfo {author} {\bibfnamefont {I.}~\bibnamefont {Poberaj}},\ }\bibfield  {title} {\bibinfo {title} {Fast and precise {Laguerre-Gaussian} beam steering with acousto-optic deflectors},\ }\href {https://doi.org/10.1364/AO.52.006506} {\bibfield  {journal} {\bibinfo  {journal} {Appl. Opt.}\ }\textbf {\bibinfo {volume} {52}},\ \bibinfo {pages} {6506} (\bibinfo {year} {2013})}\BibitemShut {NoStop}%
\bibitem [{\citenamefont {Fontaine}\ \emph {et~al.}(2021)\citenamefont {Fontaine}, \citenamefont {Ryf}, \citenamefont {Chen}, \citenamefont {Neilson}, \citenamefont {Kim},\ and\ \citenamefont {Carpenter}}]{Fontaine2019}%
  \BibitemOpen
  \bibfield  {author} {\bibinfo {author} {\bibfnamefont {N.~K.}\ \bibnamefont {Fontaine}}, \bibinfo {author} {\bibfnamefont {R.}~\bibnamefont {Ryf}}, \bibinfo {author} {\bibfnamefont {H.}~\bibnamefont {Chen}}, \bibinfo {author} {\bibfnamefont {D.~T.}\ \bibnamefont {Neilson}}, \bibinfo {author} {\bibfnamefont {K.}~\bibnamefont {Kim}},\ and\ \bibinfo {author} {\bibfnamefont {J.}~\bibnamefont {Carpenter}},\ }\bibfield  {title} {\bibinfo {title} {{Laguerre-Gaussian} mode sorter},\ }\href {https://doi.org/doi.org/10.1038/s41467-019-09840-4} {\bibfield  {journal} {\bibinfo  {journal} {Nature Communications}\ }\textbf {\bibinfo {volume} {10}},\ \bibinfo {pages} {1} (\bibinfo {year} {2021})}\BibitemShut {NoStop}%
\bibitem [{\citenamefont {Droop}\ \emph {et~al.}(2021)\citenamefont {Droop}, \citenamefont {Asché}, \citenamefont {Otte},\ and\ \citenamefont {Denz}}]{Droop2021}%
  \BibitemOpen
  \bibfield  {author} {\bibinfo {author} {\bibfnamefont {R.}~\bibnamefont {Droop}}, \bibinfo {author} {\bibfnamefont {E.}~\bibnamefont {Asché}}, \bibinfo {author} {\bibfnamefont {E.}~\bibnamefont {Otte}},\ and\ \bibinfo {author} {\bibfnamefont {C.}~\bibnamefont {Denz}},\ }\bibfield  {title} {\bibinfo {title} {Shaping light in 3d space by counter-propagation},\ }\href {https://doi.org/doi.org/10.1038/s41598-021-97313-4} {\bibfield  {journal} {\bibinfo  {journal} {Scientific Reports}\ }\textbf {\bibinfo {volume} {11}},\ \bibinfo {pages} {1} (\bibinfo {year} {2021})}\BibitemShut {NoStop}%
\bibitem [{\citenamefont {Parisi}\ \emph {et~al.}(2014)\citenamefont {Parisi}, \citenamefont {Mari}, \citenamefont {Spinello}, \citenamefont {Romanato},\ and\ \citenamefont {Tamburini}}]{Parisi2014}%
  \BibitemOpen
  \bibfield  {author} {\bibinfo {author} {\bibfnamefont {G.}~\bibnamefont {Parisi}}, \bibinfo {author} {\bibfnamefont {E.}~\bibnamefont {Mari}}, \bibinfo {author} {\bibfnamefont {F.}~\bibnamefont {Spinello}}, \bibinfo {author} {\bibfnamefont {F.}~\bibnamefont {Romanato}},\ and\ \bibinfo {author} {\bibfnamefont {F.}~\bibnamefont {Tamburini}},\ }\bibfield  {title} {\bibinfo {title} {Manipulating intensity and phase distribution of composite {Laguerre-Gaussian} beams},\ }\href {https://doi.org/10.1364/OE.22.017135} {\bibfield  {journal} {\bibinfo  {journal} {Opt. Express}\ }\textbf {\bibinfo {volume} {22}},\ \bibinfo {pages} {17135} (\bibinfo {year} {2014})}\BibitemShut {NoStop}%
\bibitem [{\citenamefont {Volyar}\ \emph {et~al.}(2022)\citenamefont {Volyar}, \citenamefont {Abramochkin}, \citenamefont {Akimova},\ and\ \citenamefont {Bretsko}}]{Volyar2022}%
  \BibitemOpen
  \bibfield  {author} {\bibinfo {author} {\bibfnamefont {A.}~\bibnamefont {Volyar}}, \bibinfo {author} {\bibfnamefont {E.}~\bibnamefont {Abramochkin}}, \bibinfo {author} {\bibfnamefont {Y.}~\bibnamefont {Akimova}},\ and\ \bibinfo {author} {\bibfnamefont {M.}~\bibnamefont {Bretsko}},\ }\bibfield  {title} {\bibinfo {title} {Control of the orbital angular momentum via radial numbers of structured {Laguerre--Gaussian} beams},\ }\href {https://doi.org/10.1364/OL.459404} {\bibfield  {journal} {\bibinfo  {journal} {Opt. Lett.}\ }\textbf {\bibinfo {volume} {47}},\ \bibinfo {pages} {2402} (\bibinfo {year} {2022})}\BibitemShut {NoStop}%
\bibitem [{\citenamefont {{Morales Rodr\'iguez}}\ \emph {et~al.}(2024)\citenamefont {{Morales Rodr\'iguez}}, \citenamefont {Maga{\~n}a-Loaiza}, \citenamefont {Perez-Garcia}, \citenamefont {Nieto}, \citenamefont {{Marroqu{\'i}n Guti{\'e}rrez}},\ and\ \citenamefont {Rodr{\'i}guez-Lara}}]{Morales2024a}%
  \BibitemOpen
  \bibfield  {author} {\bibinfo {author} {\bibfnamefont {M.~P.}\ \bibnamefont {{Morales Rodr\'iguez}}}, \bibinfo {author} {\bibfnamefont {O.~S.}\ \bibnamefont {Maga{\~n}a-Loaiza}}, \bibinfo {author} {\bibfnamefont {B.}~\bibnamefont {Perez-Garcia}}, \bibinfo {author} {\bibfnamefont {L.~M.}\ \bibnamefont {Nieto}}, \bibinfo {author} {\bibfnamefont {F.}~\bibnamefont {{Marroqu{\'i}n Guti{\'e}rrez}}},\ and\ \bibinfo {author} {\bibfnamefont {B.~M.}\ \bibnamefont {Rodr{\'i}guez-Lara}},\ }\bibfield  {title} {\bibinfo {title} {Coherent states of {Laguerre-Gauss} modes},\ }\href {https://doi.org/10.1364/OL.511439} {\bibfield  {journal} {\bibinfo  {journal} {Opt. Lett.}\ }\textbf {\bibinfo {volume} {49}},\ \bibinfo {pages} {1489} (\bibinfo {year} {2024})}\BibitemShut {NoStop}%
\bibitem [{\citenamefont {{Morales Rodr{\'i}guez}}\ \emph {et~al.}(2024)\citenamefont {{Morales Rodr{\'i}guez}}, \citenamefont {{Garc{\'i}a Herrera}}, \citenamefont {{Maga{\~n}a-Loaiza}}, \citenamefont {Perez-Garcia}, \citenamefont {{Marroqu{\'i}n Gut{\'i}errez}},\ and\ \citenamefont {Rodr{\'i}guez-Lara}}]{Morales2024b}%
  \BibitemOpen
  \bibfield  {author} {\bibinfo {author} {\bibfnamefont {M.~P.}\ \bibnamefont {{Morales Rodr{\'i}guez}}}, \bibinfo {author} {\bibfnamefont {E.}~\bibnamefont {{Garc{\'i}a Herrera}}}, \bibinfo {author} {\bibfnamefont {O.~S.}\ \bibnamefont {{Maga{\~n}a-Loaiza}}}, \bibinfo {author} {\bibfnamefont {B.}~\bibnamefont {Perez-Garcia}}, \bibinfo {author} {\bibfnamefont {F.}~\bibnamefont {{Marroqu{\'i}n Gut{\'i}errez}}},\ and\ \bibinfo {author} {\bibfnamefont {B.~M.}\ \bibnamefont {Rodr{\'i}guez-Lara}},\ }\bibfield  {title} {\bibinfo {title} {Spatial light mode analogues of generalized quantum coherent states},\ }\href {https://doi.org/10.1103/PhysRevA.110.033523} {\bibfield  {journal} {\bibinfo  {journal} {Phys. Rev. A}\ }\textbf {\bibinfo {volume} {110}},\ \bibinfo {pages} {033523} (\bibinfo {year} {2024})}\BibitemShut {NoStop}%
\bibitem [{\citenamefont {{Collado Hern\'andez}}\ \emph {et~al.}(2024)\citenamefont {{Collado Hern\'andez}}, \citenamefont {{Marroqu{\'i}n Guti{\'e}rrez}},\ and\ \citenamefont {Rodr{\'i}guez-Lara}}]{Collado2024}%
  \BibitemOpen
  \bibfield  {author} {\bibinfo {author} {\bibfnamefont {A.}~\bibnamefont {{Collado Hern\'andez}}}, \bibinfo {author} {\bibfnamefont {F.}~\bibnamefont {{Marroqu{\'i}n Guti{\'e}rrez}}},\ and\ \bibinfo {author} {\bibfnamefont {B.~M.}\ \bibnamefont {Rodr{\'i}guez-Lara}},\ }\bibfield  {title} {\bibinfo {title} {Harmonic motion modes in parabolic {GRIN} fibers},\ }\href {https://doi.org/10.1364/OPTCON.525575} {\bibfield  {journal} {\bibinfo  {journal} {Optics Continuum}\ }\textbf {\bibinfo {volume} {3}},\ \bibinfo {pages} {1025} (\bibinfo {year} {2024})}\BibitemShut {NoStop}%
\bibitem [{\citenamefont {Bandres}\ and\ \citenamefont {Guti\'{e}rrez-Vega}(2004)}]{Bandres2004}%
  \BibitemOpen
  \bibfield  {author} {\bibinfo {author} {\bibfnamefont {M.~A.}\ \bibnamefont {Bandres}}\ and\ \bibinfo {author} {\bibfnamefont {J.~C.}\ \bibnamefont {Guti\'{e}rrez-Vega}},\ }\bibfield  {title} {\bibinfo {title} {{Ince--Gaussian} modes of the paraxial wave equation and stable resonators},\ }\href {https://doi.org/10.1364/JOSAA.21.000873} {\bibfield  {journal} {\bibinfo  {journal} {J. Opt. Soc. Am. A}\ }\textbf {\bibinfo {volume} {21}},\ \bibinfo {pages} {873} (\bibinfo {year} {2004})}\BibitemShut {NoStop}%
\bibitem [{\citenamefont {Abramochkin}\ and\ \citenamefont {Volostnikov}(2004)}]{Abramochkin2004}%
  \BibitemOpen
  \bibfield  {author} {\bibinfo {author} {\bibfnamefont {E.~G.}\ \bibnamefont {Abramochkin}}\ and\ \bibinfo {author} {\bibfnamefont {V.~G.}\ \bibnamefont {Volostnikov}},\ }\bibfield  {title} {\bibinfo {title} {Generalized {Gaussian} beams},\ }\href {https://doi.org/10.1088/1464-4258/6/5/001} {\bibfield  {journal} {\bibinfo  {journal} {Journal of Optics A: Pure and Applied Optics}\ }\textbf {\bibinfo {volume} {6}},\ \bibinfo {pages} {S157} (\bibinfo {year} {2004})}\BibitemShut {NoStop}%
\bibitem [{\citenamefont {Abramochkin}\ and\ \citenamefont {Volostnikov}(1991)}]{Abramochkin1991}%
  \BibitemOpen
  \bibfield  {author} {\bibinfo {author} {\bibfnamefont {E.}~\bibnamefont {Abramochkin}}\ and\ \bibinfo {author} {\bibfnamefont {V.}~\bibnamefont {Volostnikov}},\ }\bibfield  {title} {\bibinfo {title} {Beam transformations and nontransformed beams},\ }\href {https://doi.org/https://doi.org/10.1016/0030-4018(91)90534-K} {\bibfield  {journal} {\bibinfo  {journal} {Optics Communications}\ }\textbf {\bibinfo {volume} {83}},\ \bibinfo {pages} {123} (\bibinfo {year} {1991})}\BibitemShut {NoStop}%
\bibitem [{\citenamefont {Wang}\ \emph {et~al.}(2016)\citenamefont {Wang}, \citenamefont {Chen}, \citenamefont {Zhang}, \citenamefont {Chen},\ and\ \citenamefont {Yu}}]{Wang2016}%
  \BibitemOpen
  \bibfield  {author} {\bibinfo {author} {\bibfnamefont {Y.}~\bibnamefont {Wang}}, \bibinfo {author} {\bibfnamefont {Y.}~\bibnamefont {Chen}}, \bibinfo {author} {\bibfnamefont {Y.}~\bibnamefont {Zhang}}, \bibinfo {author} {\bibfnamefont {H.}~\bibnamefont {Chen}},\ and\ \bibinfo {author} {\bibfnamefont {S.}~\bibnamefont {Yu}},\ }\bibfield  {title} {\bibinfo {title} {Generalised {Hermite–Gaussian} beams and mode transformations},\ }\href {https://doi.org/10.1088/2040-8978/18/5/055001} {\bibfield  {journal} {\bibinfo  {journal} {Journal of Optics}\ }\textbf {\bibinfo {volume} {18}},\ \bibinfo {pages} {055001} (\bibinfo {year} {2016})}\BibitemShut {NoStop}%
\bibitem [{\citenamefont {Matsumoto}\ \emph {et~al.}(2008)\citenamefont {Matsumoto}, \citenamefont {Ando}, \citenamefont {Inoue}, \citenamefont {Ohtake}, \citenamefont {Fukuchi},\ and\ \citenamefont {Hara}}]{Matsumoto2008}%
  \BibitemOpen
  \bibfield  {author} {\bibinfo {author} {\bibfnamefont {N.}~\bibnamefont {Matsumoto}}, \bibinfo {author} {\bibfnamefont {T.}~\bibnamefont {Ando}}, \bibinfo {author} {\bibfnamefont {T.}~\bibnamefont {Inoue}}, \bibinfo {author} {\bibfnamefont {Y.}~\bibnamefont {Ohtake}}, \bibinfo {author} {\bibfnamefont {N.}~\bibnamefont {Fukuchi}},\ and\ \bibinfo {author} {\bibfnamefont {T.}~\bibnamefont {Hara}},\ }\bibfield  {title} {\bibinfo {title} {Generation of high-quality higher-order {Laguerre-Gaussian} beams using liquid-crystal-on-silicon spatial light modulators},\ }\href {https://doi.org/10.1364/JOSAA.25.001642} {\bibfield  {journal} {\bibinfo  {journal} {J. Opt. Soc. Am. A}\ }\textbf {\bibinfo {volume} {25}},\ \bibinfo {pages} {1642} (\bibinfo {year} {2008})}\BibitemShut {NoStop}%
\bibitem [{\citenamefont {Aguirre-Olivas}\ \emph {et~al.}(2015)\citenamefont {Aguirre-Olivas}, \citenamefont {Mellado-Villase{\~n}or}, \citenamefont {{S\'anchez-de-la-Llave}},\ and\ \citenamefont {Arriz{\'o}n}}]{Aguirre2015}%
  \BibitemOpen
  \bibfield  {author} {\bibinfo {author} {\bibfnamefont {D.}~\bibnamefont {Aguirre-Olivas}}, \bibinfo {author} {\bibfnamefont {G.}~\bibnamefont {Mellado-Villase{\~n}or}}, \bibinfo {author} {\bibfnamefont {D.}~\bibnamefont {{S\'anchez-de-la-Llave}}},\ and\ \bibinfo {author} {\bibfnamefont {V.}~\bibnamefont {Arriz{\'o}n}},\ }\bibfield  {title} {\bibinfo {title} {Efficient generation of {Hermite-Gauss} and {Ince-Gauss} beams through kinoform phase elements},\ }\href {https://doi.org/10.1364/AO.54.008444} {\bibfield  {journal} {\bibinfo  {journal} {Appl. Opt.}\ }\textbf {\bibinfo {volume} {54}},\ \bibinfo {pages} {8444} (\bibinfo {year} {2015})}\BibitemShut {NoStop}%
\bibitem [{\citenamefont {Arriz{\'o}n}\ \emph {et~al.}(2007)\citenamefont {Arriz{\'o}n}, \citenamefont {Ruiz}, \citenamefont {Carrada},\ and\ \citenamefont {Gonz{\'a}lez}}]{Arrizon2007}%
  \BibitemOpen
  \bibfield  {author} {\bibinfo {author} {\bibfnamefont {V.}~\bibnamefont {Arriz{\'o}n}}, \bibinfo {author} {\bibfnamefont {U.}~\bibnamefont {Ruiz}}, \bibinfo {author} {\bibfnamefont {R.}~\bibnamefont {Carrada}},\ and\ \bibinfo {author} {\bibfnamefont {L.~A.}\ \bibnamefont {Gonz{\'a}lez}},\ }\bibfield  {title} {\bibinfo {title} {Pixelated phase computer holograms for the accurate encoding of scalar complex fields},\ }\href {https://doi.org/10.1364/JOSAA.24.003500} {\bibfield  {journal} {\bibinfo  {journal} {J. Opt. Soc. Am. A}\ }\textbf {\bibinfo {volume} {24}},\ \bibinfo {pages} {3500} (\bibinfo {year} {2007})}\BibitemShut {NoStop}%
\end{thebibliography}

%

\end{document}